\documentclass[titlepage,twoside,12pt]{article}
\usepackage{amssymb}
\usepackage{amsfonts}
\begin{document}

{\Large \bf A reply to P W Shor} \\ \\

{\bf Elem\'{e}r E ~Rosinger} \\
{\small \it Department of Mathematics \\ and Applied Mathematics} \\
{\small \it University of Pretoria} \\
{\small \it Pretoria} \\
{\small \it 0002 South Africa} \\
{\small \it eerosinger@hotmail.com} \\ \\

{\bf Abstract} \\

Two additional reasons are suggested for the seeming lack of progress in producing quantum
algorithms. \\ \\

In his recent paper, [1], P W Shor brings several arguments related to the somewhat unexpected
fact that, unlike in other branches of Quantum Information Processing, there has during the
last decade or so not been a more spectacular or remarkable progress with respect to new
quantum algorithms. And in this regard, he suggests that, quite likely, one should focus on
those problems which in their complexity happen to fall between the P and the NP-hard ones.
This suggestion is, of course, made upon the assumption that quantum computers will never be
able to solve in polynomial time NP-complete problems, and it also faces the difficulty that
there are not many known problems in that intermediary category. Later, Shor makes the
additional suggestion that quantum algorithms may also prove useful in solving P complexity
problems faster than the known algorithms. \\

Here, related to the above, we make two comments. \\

First, one should not forget that when going from usual algorithms to quantum algorithms one
is inevitably subjected to a process described by the classical "you lose some, you win some".
Indeed, in usual algorithms, the typical gates are {\it not} reversible, while on the contrary,
in quantum computers they are. On the other hand, quantum algorithms allow several gates which
are highly convenient, yet are not available in usual algorithms. As a consequence of such a
"you lose some, you win some" situation, one cannot simply transcribe a usual algorithm into a
quantum one. Instead, when given a problem which already has a usual algorithm, one quite
likely has to reinvent a whole new algorithm, this time in quantum terms. \\

Second, we do not yet have effectively functioning quantum computers of any practically useful,
let alone, important size. Therefore, there is no particularly tempting incentive to sit down
and get into the well known drudgery of inventing algorithms, and in this case, quantum ones,
thus doing it from the scratch, plus in that situation of "you lose some, you win some". A
good example in this regard is the way algorithms involving massive parallel computing started
to be developed in the 1980s, and not before that. Indeed, the respective problem of the von
Neumann computer architecture had been there ever since their inception in the 1940s, namely,
that such an architecture can allow only one single computation at each moment. This is a
situation like that in which two immense armies face each other, but at each moment only one
single soldier from each of them can fight the opponent. In the 1980s, however, computer
technology developed enough in order to be able to perform sufficiently massive parallel
computation. And then, when large numbers of "soldiers" from the respective "armies" could
suddenly "fight" one another, there was a significant incentive to develop new algorithms able
to do the same. \\

Needless to say, we are still quite at the beginning of the venture of involving quantum
processes in information technology. And then, it is not so easy to have a clear enough view
of what happens, what does not happen, and even less so of the respective possible
reasons. \\ \\

[1] Peter W Shor : Progress in Quantum Algorithms. Quantum Information Processing, Vol. 3,
No. 1-5, October 2004, 5-13

\end{document}